\newdimen\sb \def\md{\sb=.01em \ifmmode $\rlap{.}$'$\kern -\sb$
                               \else \rlap{.}$'$\kern -\sb\fi}
\newdimen\sa \def\sd{\sa=.1em  \ifmmode $\rlap{.}$''$\kern -\sa$
                               \else \rlap{.}$''$\kern -\sa\fi}
\message{STBASIC.TEX TeX Macro Library}
\message{ }





\def\beginrefs{\begingroup\parindent=0pt\frenchspacing
   \parskip=1pt plus 1pt minus 1pt\interlinepenalty=1000\pretolerance=10000
   \hyphenpenalty=10000\everypar={\hangindent=0.42in}       
  \def\aa##1{{\it Astr.~Ap., \bf ##1}}
  \def\aasup##1{{\it Astr.~Ap.~Suppl., \bf ##1}}
  \def\aasupp##1{{\it Astr.~Ap.~Suppl., \bf ##1}}
  \def\aj##1{{\it A.~J., \bf ##1}}
  \def\annrev##1{{\it Ann.~Rev.\ Astr.~Ap., \bf ##1}}     
  \def\araa##1{{\it Ann.~Rev.\ Astr.~Ap., \bf ##1}}     
  \def\apj##1{{\it Ap.~J., \bf ##1}}     
  \def\apjl##1{{\it Ap.~J. (Letters), \bf ##1}}
  \def\apjlett##1{{\it Ap.~J. (Letters), \bf ##1}}
  \def\apjlet##1{{\it Ap.~J. (Letters), \bf ##1}}
  \def\apjsup##1{{\it Ap.~J.~Suppl., \bf ##1}}
  \def\apjsupp##1{{\it Ap.~J.~Suppl., \bf ##1}}
  \def\baas##1{{\it Bull.~A.A.S., \bf ##1}}
  \def\ban##1{{\it B.A.N., \bf ##1}}
  \def\ibvs##1{{\it Inf. Bull. Var. Stars}, No.~##1}
  \def\mn##1{{\it M.N.R.A.S., \bf ##1}}
  \def\mnras##1{{\it M.N.R.A.S., \bf ##1}}
  \def\pasp##1{{\it Pub.~A.S.P., \bf ##1}}
  \def\ajpasp##1{{\it Pub.~A.S.P., \bf ##1}}
  \def\nat##1{{\it Nature, \bf ##1}}
  \def\nature##1{{\it Nature, \bf ##1}}}

\def\endrefs{\endgroup}



\def\df{\leaders\hbox to 0.6em{\hss.}\hfill}


\def\section#1{\bigbreak\medskip\centerline{#1}\par\nobreak\medskip\markpage}

\def\subsection#1#2{\bigbreak\noindent{\bf#1\hskip 0.9em\relax#2}\par
   \nobreak\medskip\markpage}

\def\subsubsection#1#2{\medbreak\noindent{\sl#1\hskip 0.60em\relax#2}\par
   \nobreak\medskip\markpage}

\def\today{\advance\year by -1900 
   \number\month/\number\day/\number\year}
\def\yearmonthday{\number\year\space
   \ifcase\month\or January\or February\or March\or April\or May\or June\or
   July\or August\or September\or October\or November\or December\fi
   \space\number\day}

\newcount\num

\def\nextnum{\global\advance \num by 1 \number\num}
\def\nextitem{\leavevmode
   \hbox{\ifnum\num>8 \kern-0.43em\fi \nextnum.\kern0.60em}}
\def\bfnextitem{\leavevmode
   \hbox{\ifnum\num>8 \kern-0.43em\fi \bf\nextnum.\kern0.60em}}

\newcount\colnum

\def\nextcolnum{\global\advance \colnum by 1 \number\colnum}
\def\nextcolumn{\leavevmode
   \hbox{{\it \ifnum\colnum<9 \phantom{1}\fi Column \nextcolnum:}\kern0.60em}}

\newcount\fig

\def\nextfig{\global\advance \fig by 1 \number\fig}

\newcount\cap

\def\nextcap{\global\advance \cap by 1 \number\cap}

\newcount\letter

\def\nextlet{\global\advance \letter by 1
   \ifcase\letter\or A\or B\or C\or D\or E\or F\or G\or H\or I\or
   J\or K\or L\or M\or N\or O\or P\or Q\or R\or S\or T\or U\or V\or W\or X\or
   Y\or Z\fi}

\newdimen\bigindent \bigindent=3.5in
\def\letterhead{\hsize=6in\interlinepenalty=2000\parskip=6pt minus 3pt
  \pretolerance=750
  \def\topline##1{\hbox to\hsize{\hfil##1\hskip\rightskip}}
  \footline={\ifnum\pageno=1
    \hss\hbox{\vrule height 0.4in width 0pt}
    \eightrm Operated by the Association of Universities for Research in 
    Astronomy, Inc., for the National Aeronautics and Space Administration\hss
    \else\hfil\fi}
  \null
  \vskip-0.2in
  {\advance\rightskip by -0.75in
    \topline{3700 San Martin Drive}
    \topline{Baltimore, MD 21218}
    \topline{(301) 338-4718}\par}
  \vskip30pt minus 15pt
  {\leftskip=\bigindent\yearmonthday\par}}

\def\arpanetletterhead{\hsize=6in\interlinepenalty=2000\parskip=6pt minus 3pt
  \pretolerance=750
  \def\topline##1{\hbox to\hsize{\hfil##1\hskip\rightskip}}
  \footline={\ifnum\pageno=1
    \hss\hbox{\vrule height 0.4in width 0pt}
    \eightrm Operated by the Association of Universities for Research in 
    Astronomy, Inc., for the National Aeronautics and Space Administration\hss
    \else\hfil\fi}
  \null
  \vskip-0.2in\vskip-3\baselineskip
  {\advance\rightskip by -0.75in
    \topline{3700 San Martin Drive}
    \topline{Baltimore, MD 21218}
    \topline{(301) 338-4718}
    \topline{{\elevenrm BITNET:} \tt golombek@stsci}
    \topline{\elevenrm SPAN: \tt SCIVAX::GOLOMBEK}
    \topline{{\elevenrm ARPANET:} \tt golombek@scivax.arpa}\par}
  \vskip30pt minus 15pt
  {\leftskip=\bigindent\yearmonthday\par}}

\def\gosbletterhead{\hsize=6in\interlinepenalty=2000\parskip=6pt minus 3pt
  \pretolerance=750
  \def\topline##1{\hbox to\hsize{\hfil##1\hskip\rightskip}}
  \footline={\ifnum\pageno=1
    \hss\hbox{\vrule height 0.4in width 0pt}
    \eightrm Operated by the Association of Universities for Research in 
    Astronomy, Inc., for the National Aeronautics and Space Administration\hss
    \else\hfil\fi}
  \null
  \vskip-0.375in
  {\advance\rightskip by -0.75in
    \topline{General Observer Support Branch}
    \topline{3700 San Martin Drive}
    \topline{Baltimore, MD 21218}
    \topline{(301) 338-4996}\par}
  \vskip30pt minus 15pt
  {\leftskip=\bigindent\yearmonthday\par}}



\def\indentleft{\advance\leftskip by 50pt\interlinepenalty=750}
\def\inndentleft{\advance\leftskip by 78pt\interlinepenalty=750}
\def\narrower{\advance\leftskip by 0.42in\advance\rightskip by 0.42in
  \interlinepenalty=750}
\def\nnarrower{\advance\leftskip by 50pt\advance\rightskip by 45pt
  \interlinepenalty=750}

\def\checkbox{\nnarrower\parindent=0pt\itemitem{\vbox{\hrule height.7pt
  \hbox{\vrule width.7pt height6pt \kern6pt \vrule width.7pt}
  \hrule height.7pt}$\,$}}  


%
%
\newcount\index \index=100
\def\markpage{\advance\index by 1 \count\index=\pageno}
\def\begintableofcontents{\begingroup
  \index=100 \frenchspacing\interlinepenalty=750
  \parskip=0.1pt plus 1pt minus 0.1pt \parindent=0.3in
  \def\dfi{\advance\index by 1 \df\number\count\index}
  \def\in{\par\hskip-0.2in\indent \hangindent2\parindent \textindent}    
  \def\inin{\par\hskip0.32in\indent \hangindent3\parindent \textindent}
  \def\ininin{\par\hskip0.95in\indent \hangindent4\parindent \textindent}}



{\obeylines\gdef\startdisplay#1
  {\catcode`\^^M=5$$#1\halign\bgroup\indent##\hfil&&\qquad##\hfil\cr}}
\outer\def\enddisplay{\crcr\egroup$$}

\chardef\other=12

{\obeyspaces\gdef {\ }} 

  \font\twentyfourrm=cmr10 scaled 2488
  \font\twentyfouri=cmmi10 scaled 2074   
  \font\twentyfoursy=cmsy10 scaled 2074
  \font\twentyrm=cmr10 scaled 2074      
  \font\twentyi=cmmi10 scaled 2074   
  \font\twentysy=cmsy10 scaled 2074
  \font\eighteenrm=cmr10 scaled 1728
  \font\eighteeni=cmmi10 scaled 1728 \font\eighteensy=cmsy10 scaled 1728
  \font\fourteenrm=cmr10 scaled 1440
  \font\fourteeni=cmmi10 scaled 1440 \font\fourteensy=cmsy10 scaled 1440
  \font\twelverm=cmr12
                
  \font\twelvei=cmmi12               \font\twelvesy=cmsy10 scaled 1200
  \font\elevenrm=cmr10 scaled 1095
    
  \font\eleveni=cmmi10 scaled 1095   \font\elevensy=cmsy10 scaled 1095
  \font\tenrm=cmr10
                   
  \font\teni=cmmi10  \font\tensy=cmsy10  
  \font\ninerm=cmr9

  \font\ninei=cmmi9                  \font\ninesy=cmsy9
  \font\eightrm=cmr8
  \font\seveni=cmmi7 \font\sevensy=cmsy7

\def\commonstuff{
  \parindent=0.42in       
  \def\skipline{\vskip\baselineskip}
  \hyphenpenalty=200\pretolerance=300\tolerance=600 
  \interlinepenalty=100\clubpenalty=500\widowpenalty=500
  \nonfrenchspacing\singlespace\rm}

\def\twelvepoint{
  \font\bf=cmbx12
  \font\it=cmti12
  \font\sl=cmsl12
  \font\tb=cmtt10 scaled 1200 
  \font\tt=cmtt8 scaled 1440
  \textfont0=\twelverm \scriptfont0=\tenrm     
    \scriptscriptfont0=\sevenrm                 
  \def\rm{\fam0 \twelverm}   
  \textfont1=\twelvei  \scriptfont1=\teni  
    \scriptscriptfont1=\seveni                  
  \def\mit{\fam1 } \def\oldstyle{\fam1 \twelvei}
  \textfont2=\twelvesy \scriptfont2=\tensy 
    \scriptscriptfont2=\sevensy                 
  \def\singlespace{\baselineskip=13.5pt\lineskiplimit=-5pt
    \lineskip=0pt
    \parskip=1.25pt plus 1.5pt minus 0.25pt}  
  \def\oneandahalfspace{\baselineskip=18pt\parskip=0pt plus 1pt}
  \def\doublespace{\baselineskip=24pt\parskip=0pt plus 0.5pt}
  \footline={\ifnum\pageno=1 \hfil
             \else\hss\twelverm-- \folio\ --\hss\fi} 
  \def\pagenumbers{\footline={\hss\twelverm-- \folio\ --\hss}}  
  \def\romanpagenumbers{\footline={\hss\twelverm-- \romannumeral\folio\ --\hss}}
  \commonstuff}

\def\tenpoint{
  \font\it=cmti10
  \font\sl=cmsl10
  \font\bf=cmb10
  \textfont0=\tenrm \scriptfont0=\sevenrm     
    \scriptscriptfont0=\fiverm                 
  \def\rm{\fam0 \tenrm}   
  \textfont1=\teni  \scriptfont1=\seveni  
    \scriptscriptfont1=\fivei                  
  \def\mit{\fam1 } \def\oldstyle{\fam1 \teni}
  \textfont2=\tensy \scriptfont2=\sevensy 
    \scriptscriptfont2=\fivesy                 
  \def\singlespace{\baselineskip=12pt\lineskiplimit=0pt
    \lineskip=-0.5mm       
    \parskip=2pt plus 1pt minus 1pt}  
  \footline={\ifnum\pageno=1 \hfil
             \else\hss\tenrm-- \folio\ --\hss\fi} 
  \def\oneandahalfspace{\baselineskip=18pt\parskip=0pt plus 1pt}
  \def\doublespace{\baselineskip=24pt\parskip=0pt plus 1 pt}
  \def\pagenumbers{\footline={\hss\tenrm-- \folio\ --\hss}}  
  \def\romanpagenumbers{\footline={\hss\tenrm-- \romannumeral\folio\ --\hss}}
  \commonstuff}

\def\elevenpoint{
  \font\it=cmti10 scaled 1095
  \font\sl=cmsl10 scaled 1095
  \font\bf=cmb10 scaled 1095 
  \font\tt=cmtt10 scaled 1095
  \textfont0=\elevenrm \scriptfont0=\tenrm     
    \scriptscriptfont0=\ninerm                 
  \def\rm{\fam0 \elevenrm}   
  \textfont1=\eleveni  \scriptfont1=\teni  
    \scriptscriptfont1=\ninei                  
  \def\mit{\fam1 } \def\oldstyle{\fam1 \eleveni}
  \textfont2=\elevensy \scriptfont2=\tensy 
    \scriptscriptfont2=\ninesy                 
  \def\singlespace{\baselineskip=13pt\lineskiplimit=-5pt
    \lineskip=0mm       
    \parskip=2pt plus 1pt minus 1pt}  
  \footline={\ifnum\pageno=1 \hfil
             \else\hss\elevenrm-- \folio\ --\hss\fi} 
  \def\oneandahalfspace{\baselineskip=19pt\parskip=0pt plus 1pt}
  \def\doublespace{\baselineskip=26pt\parskip=0pt plus 1 pt}
  \def\pagenumbers{\footline={\hss\elevenrm-- \folio\ --\hss}}  
  \def\romanpagenumbers{\footline={\hss\tenrm-- \romannumeral\folio\ --\hss}}
  \commonstuff}

\def\eighteenpoint{           
  \font\bf=cmbx10 scaled 1728
  \font\it=cmti10 scaled 1728
  \font\sl=cmsl10 scaled 1728
  \font\tb=cmtt10 scaled 1728
  \font\tt=cmtt10 scaled 1728
  \textfont0=\eighteenrm \scriptfont0=\fourteenrm
    \scriptscriptfont0=\twelverm                 
  \def\rm{\fam0 \eighteenrm}   
  \textfont1=\eighteeni  \scriptfont1=\fourteeni  
    \scriptscriptfont1=\twelvei                  
  \def\mit{\fam1 } \def\oldstyle{\fam1 \eighteeni}
  \textfont2=\eighteensy \scriptfont2=\fourteensy 
    \scriptscriptfont2=\twelvesy                 
  \def\singlespace{\baselineskip=21pt\lineskiplimit=-5pt
    \lineskip=0pt
    \parskip=4pt plus 1pt minus 1pt}  
  \def\oneandahalfspace{\baselineskip=30pt\parskip=0pt plus 1pt}
  \def\doublespace{\baselineskip=40pt\parskip=0pt plus 1pt}
  \footline={\ifnum\pageno=1 \hfil
             \else\hss\eighteenrm-- \folio\ --\hss\fi} 
  \def\pagenumbers{\footline={\hss\eighteenrm-- \folio\ --\hss}}  
  \commonstuff}

\def\twentypoint{
  \font\bf=cmbx10 scaled 2074
  \font\it=cmti10 scaled 2074
  \font\sl=cmsl10 scaled 2074
  \font\tb=cmtt10 scaled 2074
  \font\tt=cmtt10 scaled 2074
  \textfont0=\twentyrm \scriptfont0=\eighteenrm     
    \scriptscriptfont0=\fourteenrm                 
  \def\rm{\fam0 \twentyrm}   
  \textfont1=\twentyi  \scriptfont1=\eighteeni  
    \scriptscriptfont1=\fourteeni                  
  \def\mit{\fam1 } \def\oldstyle{\fam1 \twentyi}
  \textfont2=\twentysy \scriptfont2=\eighteensy 
    \scriptscriptfont2=\fourteensy                 
  \def\singlespace{\baselineskip=24pt\lineskiplimit=-5pt
    \lineskip=0pt
    \parskip=5pt plus 1.5pt minus 1.5pt}  
  \def\oneandahalfspace{\baselineskip=33pt\parskip=0pt plus 1pt}
  \def\doublespace{\baselineskip=44pt\parskip=0pt plus 0.5pt}
  \footline={\ifnum\pageno=1 \hfil
             \else\hss\twentyrm-- \folio\ --\hss\fi} 
  \def\pagenumbers{\footline={\hss\twentyrm-- \folio\ --\hss}}  
  \def\romanpagenumbers{\footline={\hss\twentyrm-- \romannumeral\folio\ --\hss}}
  \commonstuff}

\def\twentyfourpoint{
  \font\bf=cmbx10 scaled 2488
  \font\it=cmti10 scaled 2488
  \font\sl=cmsl10 scaled 2488
  \font\tb=cmtt10 scaled 2488
  \font\tt=cmtt10 scaled 2488
  \textfont0=\twentyfourrm \scriptfont0=\twentyrm     
    \scriptscriptfont0=\eighteenrm                 
  \def\rm{\fam0 \twentyfourrm}   
  \textfont1=\twentyfouri  \scriptfont1=\twentyi  
    \scriptscriptfont1=\eighteeni                  
  \def\mit{\fam1 } \def\oldstyle{\fam1 \twentyfouri}
  \textfont2=\twentyfoursy \scriptfont2=\twentysy 
    \scriptscriptfont2=\eighteensy                 
  \def\singlespace{\baselineskip=28pt\lineskiplimit=-5pt
    \lineskip=0pt
    \parskip=5pt plus 1.5pt minus 1.5pt}  
  \def\oneandahalfspace{\baselineskip=42pt\parskip=0pt plus 1pt}
  \def\doublespace{\baselineskip=56pt\parskip=0pt plus 0.5pt}
  \footline={\ifnum\pageno=1 \hfil
             \else\hss\twentyfourrm-- \folio\ --\hss\fi} 
  \def\pagenumbers{\footline={\hss\twentyfourrm-- \folio\ --\hss}}  
  \def\romanpagenumbers{\footline={\hss\twentyfourrm-- \romannumeral\folio\ --\hss}}
  \commonstuff}

\def\spose#1{\hbox to 0pt{#1\hss}}
\def\lta{\mathrel{\spose{\lower 3pt\hbox{$\mathchar"218$}}
     \raise 2.0pt\hbox{$\mathchar"13C$}}}
\def\gta{\mathrel{\spose{\lower 3pt\hbox{$\mathchar"218$}}
     \raise 2.0pt\hbox{$\mathchar"13E$}}}

\def\ni{\noindent}
\def\in{\indent}
\def\inin{\in{\in}
\def\ininin{\inin{\in}}}

\twelvepoint
\doublespace
\raggedbottom
\null\vskip.5in
\vskip.4in
\centerline{\bf  
Dwarf galaxies in four rich clusters with $0.02 < z < 0.14$} 
\vskip.4in
\centerline{Neil Trentham }
\smallskip
\medskip
\centerline{Institute for Astronomy, University of Hawaii}
\vskip 4pt
\centerline{2680 Woodlawn Drive, Honolulu HI 96822, U.~S.~A.}
\vskip 4pt
\centerline{email : nat@newton.ifa.hawaii.edu}
\vskip.3in
\vskip.3in
\centerline{Submitted to $MNRAS$   } 
\vskip.3in
\vskip 50pt

\centerline{\bf ABSTRACT }
\bigskip
\noindent
Deep measurements are presented of four rich clusters of galaxies:
Abell 1367 ($z=0.022$), Abell 2199
($z=0.030$), Abell 1795 ($z=0.063$), and Abell 1146 ($z=0.141$).
All clusters have an excess of galaxies at faint magnitudes above
blank sky fields. 
We correct for background contamination and
measure the luminosity function of these galaxies in each
cluster, and then combine these luminosity functions to get better
statistics.  
The resultant combined
luminosity function is rising at faint magnitudes,
with a logarithmic slope $-1.5 < \alpha < -1.2$ for 
$-18 < M_B < -13$ and $-19 < M_R < -15$.  This is similar to
what has been observed independently in the Coma cluster.
The colours of  
these faint galaxies suggest that they are dwarf spheroidals.


\vfil\eject

\noindent{\bf 1 INTRODUCTION}

\noindent
Galaxies have masses which vary by at least
six orders of magnitude.
Yet they separate unambiguously
into just two distinct classes of objects in the
fundamental plane parameter correlations (see e.g.~Kormendy
1985, 1987; Binggeli 1994).
The first of these is the early-type giants and bulges of disk galaxies,
which have increasing  
surface-brightness as their total luminosity increases.
The second of these are the late-type giant and dwarf galaxies
which have decreasing surface-brightness as 
their total luminosity increases (they also have increasing dark
matter fractions as their total luminosity decreases; Kormendy 1990).
The giant galaxies have been well studied over the last fifty years,
but the properties of the dwarf galaxies population are only now
beginning to be described in detail. 

The first such 
measurements were of the Local Group dwarfs (see Hodge
1989 for a review) and of the dwarfs in nearby poor groups 
(Ferguson \& Sandage 1991) and poor clusters like
Virgo (Sandage et al.~1985) and Fornax (Ferguson 1989).   
These measurements showed that 
dwarf galaxies 
have a luminosity function (LF: $\phi (L) \propto
L^{\alpha}$) 
that is rising as $\alpha = -1.35$ 
at the faintest magnitudes probed ($M_B \sim -13$ for
$H_0 = 75$ km s$^{-1}$ Mpc$^{-1}$).
In these poor clusters and groups, overwhelmingly
the most numerous 
galaxy type at faint magnitudes
($M_B < -16$) are the dwarf spheroidal
(dSph), alternatively 
called dwarf elliptical, galaxies (see Binggeli
1987, 1994; Ferguson \& Binggeli 1994).
These galaxies have similar scaling laws to dwarf irregulars
(dIrr) 
in the fundamental plane correlations, but are redder, 
because old stellar populations contribute proportionately
more  
of the light.

Another important development has been the detection of 
a population of low-luminosity star-forming galaxies in the
field at $z>0.2$ (Broadhurst et al.~1988,
Cowie et al.~1991).  These contribute
a significant, but probably not the dominant 
(see Cowie et al.~1995,
1996), contribution to the excess blue counts observed
there.  The fate of these galaxies is not known, but they
might evolve into low surface-brightness galaxies
(McGaugh 1994, Ferguson \& McGaugh 1995), perhaps similar to
the dwarf spheroidal galaxies observed today (Kormendy \&
Bender 1994).  
 
More recently, measurements of dwarf galaxies in rich
clusters have been made.  This is now technically feasible
because
the contamination from background galaxies at faint magnitudes
has been quantified well (Driver et al.~1994, Bernstein
et al.~1995, Trentham 1997a) and can be corrected for, allowing
LFs to be computed.  Previous studies of clusters concentrated
on giant galaxies; these bright galaxies are 
sufficiently numerous relative to the background that
contamination effects are small. 
Much of the recent work 
has concentrated on clusters at $z \sim 0.2$ like
Abell 963 (Driver et al.~1994, Trentham 1997b),
Abell 665 (Wilson et al.~1997, Trentham 1997b), and
Abell 1689 (Wilson et al.~1997).  This redshift is
particularly interesting because it is where the
low-luminosity star-forming galaxies that contribute to
the excess blue counts begin to appear. 
However, in clusters this distant, measurements of dwarf
galaxies that are in the cluster can only
be made down to $M_B \sim -15$, so that they probe 
only a small part of the magnitude range where the
dwarfs dominate.  Nevertheless, it is interesting to note
that the dwarfs seen in these clusters appear to be
red (Trentham 1997b).
Fainter dwarfs have been observed in the
Coma cluster at $z=0.023$ (Secker \& Harris 1996,
Trentham 1997c) and are very numerous there,
with $-1.4 < \alpha < -1.7$ fainter than $M_R = -16$.
These dwarf galaxies have $1.3 < B-R < 1.5$, and are
probably dSphs. 

In this work, we present similar measurements for four
more clusters, with $0.022 < z < 0.14$.
This will permit us to see if the results found in Coma
are a universal feature of rich clusters (Coma is
anomalous in its X-ray properties as it has no
cooling flow $-$ Hughes et al.~1988 $-$ so that
galaxy evolution might well proceed
differently there to in
other rich clusters), and if
the LF or the colours of the dwarf galaxies in clusters
vary with redshift.  
 
Our sample is presented in Table 1.  There we list
the Abell (1958) richness, redshift $z$, velocity dispersion
$\sigma$, X-ray luminosity $L_x$, Galactic extinction $A_B$ along 
our line of sight, critical surface density for gravitational
lensing of distant background galaxies $\Sigma_c$, and mass
deposition rate ${\mathaccent 95{M}}$ for the cooling flow clusters.
Abell 1367 is at approximately the 
same distance as Coma, but
is not as rich, and has lower X-ray luminosity (like
Coma, it has
no cooling flow).  Its giant galaxy luminosity function has
been measured  
by Godwin \& Peach (1982) using photographic photometry.  It
is the next best example after Coma of a rich cluster at
$z \sim 0.02$ with a galaxy density high enough that we might
measure the dwarf galaxy LF. 
Abell 2199 and Abell
1795 are more distant.  Both are rich cooling flow
clusters with supergiant cD galaxies at their center.
Abell 1795 has an anomalously low 
total giant galaxy optical
luminosity given its X-ray gas mass (Arnaud et al.~1992);
the measurements here will enable us to  
tell if some of this deficiency can be made up by an excess
of dwarf galaxies.
Abell 1146 is more distant, but is extremely
rich; it is one of only seven richness 4 clusters in the
Abell catalog and is significantly the nearest of these. 
Therefore it is a good candidate for having a detectable
dwarf galaxy population.

Throughout this work we assume
$H_0 = 75$ km s$^{-1}$ Mpc$^{-1}$ and $\Omega_0 = 1$. 

\vskip 10pt

\noindent{\bf 2 OBSERVATIONS AND PHOTOMETRY} 

\vskip 5pt

\noindent{\bf 2.1 Observations and data preprocessing}

\noindent
In Figure 1, we present our images.  The observing log is presented
in Table 2, where we list the field position and 
size, total exposure time, average
extinction $<X>$ due to clouds, and seeing, for each image. 
All observations 
were taken
at the f/10 Cassegrain focus of the University of Hawaii 2.2 m telescope
on Mauna Kea 
using a thinned Tektronix 2048 $\times$ 2048 CCD 
(scale 0\sd22 pixel$^{-1}$;
field of view
7\md5 $\times$ 7\md5).
Each image was constructed from a number of shorter exposures,
of length $5-10$ minutes, dithered in order to reject cosmic rays and
bad pixels. 
These images were bias-subtracted (the CCD dark current
was negligible) and then flatfielded (the
flatfield we used was constructed using both median sky and
twilight flats) 
before
being combined. 
The reduced images are flat to better than one percent.

Instrumental magnitudes were computed from observations of several
(typically $\sim 30$ per night) standard stars 
and the photometry was converted to
the Johnson (UBV) $-$ Cousins (RI)
magnitude system of Landolt (1992).  This conversion is accurate to
about 2\%. 
Conditions were photometric when most of the images were taken; in the
instances when there was thin cirrus overhead, the images were
calibrated using shorter exposures taken under photometric conditions.  
Table 2 shows how big these corrections were.  

\vskip 5pt

\noindent{\bf 2.2 Photometry techniques}

\noindent 
We obtain total magnitudes for the galaxies we detect in our
images by measuring
isophotal magnitudes and making a correction that takes the
surface-brightness of the galaxies into account.  This method is
described in detail elsewhere (Trentham 1997a); here we list the main
steps in outline form  
and add comments that are specific to the data presented in this 
paper.

\noindent 1) We measure the rms sky noise $\sigma_{rms}$ and
the FWHM seeing $b_{\rm FWHM}$ for each image.
The sky Poisson noise is always the dominant noise source.
\vskip 1pt
\noindent 2) We then simulate galaxies of various
apparent magnitudes and exponential scale-lengths.
These simulated galaxies are then
convolved with a Gaussian seeing function of width 
$b_{\rm FWHM}$ and  
Poisson noise of rms magnitude $\sigma_{rms}$ is added. 
\vskip 1pt
\noindent 3) We then run the FOCAS detection algorithm 
(Jarvis \& Tyson 1981, Valdes 1982, 1989)
on this image to search for objects with fluxes
that are 3$\sigma_{\rm rms}$ above the sky.
For each object we measure the isophotal magnitude $m_I$ and 
first-moment light radius
$r_{1}$.
As the true magnitude $m$ of each object is known, we compute
the function $m(m_I, r_1)$, and its uncertainty  
$\sigma (m)[m_I, r_1]$.
The uncertainty comes from studying how intrinsically
identical galaxies are detected differently depending on
where they are placed in the image i.e.~depending
on the local noise.
We also determine the faintest magnitude $m_L$ at
which galaxies whose intrinsic magnitudes and scale-lengths
are equal to those of local dwarf galaxies projected to the
distance of the cluster are detected with 100\% completeness. 
This will be the faintest magnitude to which we determine
the LF in each image.  We make this cut because the
completeness is a strong function of the detailed galaxy
properties fainter than $m_L$, so trying to correct for
this effect in our data is dangerous. 
\vskip 1pt
\noindent 4) We then run the same detection algorithm
on our data, and make a catalog of all objects detected
at the 3$\sigma$ level.
For each
object, we measure $m_I$ 
and $r_{1}$. 
We also compute the aperture magnitude
$m_a$ within an aperture diameter of 3\sd0.
These magnitudes will be used to measure colours. 
Isophotal magnitudes are not used for computing colours 
because the detection isophotes are not the same in
the different bands.
A 3\sd0 
aperture is large enough that differential seeing effects between
our $B$ and $R$ images are negligible.
\vskip 1pt
\noindent 5)  
Multiple objects within a single detection isophote
are identified by searching for
multiple maxima and are split into individual objects using
the FOCAS splitting algorithm.  The algorithm is run several times
so that cases where many objects were contained in a single
isophote initially are all individually recovered. 
The quantities $m_I$ and $r_1$ are
computed for each object at each stage of the splitting.
\vskip 1pt
\noindent 6)
Objects are then classified (see Valdes 1989 for the details of the 
classification terminology) based on their morphology relative to that of
several reference PSF stars in the field.
\vskip 1pt
\noindent 7)
We then remove from the 
catalog: (i) objects whose $m (m_I, r_1) > m_L$;  
(ii) diffraction spikes of bright stars, ghost images, and chip defects
(these were identified from the FOCAS classification $-$ see 
Trentham 1997a and
Valdes 1989 for details);
(iii) spurious objects that were detected in the halos of bright stars
and galaxies where the noise is much higher than in the rest of the
image.  
The last of these  
corrections required us to look at all objects in our image that
were part of a larger object that was subsequently split into
more than three smaller objects in our original detection pass and
make a judgment by eye as to whether the faint objects we see are
real objects or enhanced noise peaks. 
Also, at this stage, a number of low surface brightness galaxies
that had been recognized as a multiple object and split into many
small objects centred on local noise peaks
were reconstructed, and the values of $m_I$ and $r_1$ computed
prior to splitting were adopted.
\vskip 1pt
\noindent 8)
After these corrections are made, we have a catalog
of objects classified as ``galaxies''
or ``stars''.  At faint magnitudes, these
classifications are unreliable because many galaxies
have apparent scale lengths
smaller than the seeing and so look like stars.   
We therefore correct for stellar contamination 
at faint
magnitudes by
computing how many faint stars we expect given the number
of bright ($m < 20$) stars 
and assuming that Galactic stellar
luminosity function 
has the slope measured 
by Jones et al.~1991.   
The stellar
contamination is small ($\sim 15$\% at the faintest
magnitudes in A1146, and much less in the other
clusters) and the
uncertainties generated by this method are also small. 
\vskip 1pt
\noindent 9)
For each object in our catalog we then use 
$m_I$ and $r_1$ 
to compute $m$ and its uncertainty $\sigma (m)$,
correct for stellar contamination as in 8),
and then bin the data in half-magnitude intervals.
The number counts 
are then computed by dividing the number of 
galaxies in each bin by the survey
area.  This survey area includes a correction for crowding, the
process by which faint galaxies go undetected because they happen to
fall within the detection isophote of a much brighter object (see 
Trentham 1997a for details of how this was done).
\vskip 1pt
\noindent 10)
The number counts were corrected for Galactic extinction using
the HI maps of Burstein \& Heiles (1982) and the colour conversions of 
Cardelli et al.~(1989).  
Corrections to the background galaxies for 
gravitational lensing
by the cluster dark matter (see Trentham 1997b) are 
$\sim 2\%$ for A1146 and much smaller for the other
clusters.
We neglect this effect, and also 
extinction from dust in the cluster (see Ferguson 1993,
Bernstein et al.~1995). 

As this stage, we have 
the number of galaxies per half-magnitude per
square degree, as a function of apparent magnitude, for each of 
our cluster fields. 
The data are presented in Section 3.1, and we describe there how
we compute the LF from them.

\vskip 10pt

\noindent{\bf 3 RESULTS AND DISCUSSION}

\vskip 5pt

\noindent{\bf 3.1 Number Counts}

\noindent
The number count $-$ magnitude relations computed as described
in Section 2.2 are presented in Figure 2.

The uncertainties are the
quadrature sum of counting statistics and uncertainties from the
isophotal corrections $\sigma (m)$.  The uncertainties from counting
statistics dominate at the bright end, and the uncertainties from 
isophotal corrections dominate at the faint end.
Also shown in Figure 2 
are the mean background counts for random 
sky fields (Trentham 1997a).   
In each case,
at progressively fainter
magnitudes the background contributes more to the total number counts.

The luminosity function is computed
by subtracting the background contribution from the number counts.
The uncertainty is computed taking into account the
field-to-field variance in the background in addition to the errors
described in the last paragraph. 
This field-to-field variance in the background counts is discussed
in detail in Trentham (1997a).
We use the numbers there, corrected here
for the differences in survey area using 
Poisson statistics.  

In converting apparent magnitudes to absolute magnitudes, we
assume distance moduli of 34.75 for A1367, 35.41 for A2199,
37.03 for A1795, and 38.83 for A1146.

\vskip 5pt

\noindent{\bf 3.2 Luminosity Functions}

\noindent
The LFs are presented in Figures 3 ($B$-band) and 4 ($R$-band).
The uncertainties are the quadrature sum of errors from
counting statistics, from uncertainties in the total magnitudes,
and from the field-to-field variance in 
the background.
The last of these dominates the
error everywhere except at
the brightest magnitudes, where counting statistics dominate.

The LF of Abell 1367 is very poorly constrained because of the large
errors following background subtraction.
This is because the galaxy density there is very low relative
to the background.  It is the poorest cluster studied here, but
no richer clusters exist at $z\sim 0.02$ except Coma, which is
studied elsewhere, and Perseus, which is close to the Galactic
plane. 
Abell 2199 has a LF that is slowly rising towards fainter
magnitudes ($\alpha \sim -1.4$) over the entire magnitude range.
This rise is apparent in both the $R$-band and $B$-band data, but
the $R$-band data has better statistics 
at the faint end ($M_R \sim -12$) following background
subtraction.  Also, the
$R$-band data probe slightly deeper ($B-R \sim 1.3$ for  
faint galaxies; see Section 3.3).
The extremely
steep LF ($\alpha = -2.2$) claimed by De Propris et al.(1995)
for the cD halo of NGC 6166 is not seen in our data.
Their estimate of the field-to-field variance in
the background ($<$1\%
probability of getting $>$1.4 times the mean counts in a
6.54 square arcminute field) 
is much smaller than our observed 
variance (16\% of getting $>$1.4 times the mean counts in
the same field at $B=24$ and more than 16\% at brighter
magnitudes; see Trentham 1997a, also Driver et al.~1994,
Bernstein et al.~1995). 
If our variance
measurement is resonable, then the errors in each
data point quoted by
De Propris et al.~must 
be underestimates.  The uncertainty
in their value of $\alpha$ then becomes large.  This is 
the most natural 
explanation of the difference in their results and ours.  
Abell 1795 has a slightly flatter LF than Abell 2199, although
we do not probe quite so faint.
We do not see enough dwarf galaxies to
explain away the
decrement in optical 
luminosity relative to X-ray gas mass described in
Section 1.
Abell 1146 also has a shallow luminosity function
for $-21 < M_R < -17$, but there is an intriguing sign of an
upturn in the faintest point in both the $R$-band ($M_R = -16$)
and $B$-band ($M_B = -15$) data.

All the LFs in each band were then combined to produce a
composite LF, which we present in Figure 5 and Tables 3 and 4.
This function falls seeply at bright magnitudes
(the cluster cD galaxies push up the LF at the
very bright end), and rises gradually towards 
faint magnitudes.

For comparison with previous deep work in other clusters and the field
and with theory, it is useful to describe the LF at each magnitude by
a slope parameter
$\alpha (M) = -2.5 { {{\rm d}\log_{10} N}\over{{\rm d}M}} - 1$.  One
method that is frequently applied is to estimate $\alpha (M)$ using
fitting functions like the Schechter (1976) function, in which $\alpha (M)$
tends to an asymptotic value $\alpha^{*}$ at faint magnitudes.
Pitfalls of this approach are reviewed in Section 2 of Trentham (1997a).
The major problems are (a) the fitting functions have no physical
significance, and (b) they are susceptible to
severe parameter coupling (e.g.~the Schechter $\alpha^{*}$ and $M^{*}$
are strongly coupled so that a value of $\alpha^{*}$ that we might
derive is strongly affected by the properties of the bright galaxies
which, as outlined in Section 1, are a separate population of stellar
systems from the dwarf galaxies.  In this work we use local linear
and polynomial fits to estimate $\alpha (M)$ at each $M$.  These fits do not
have any physical significance either, but are more useful than
Schechter fits for the present study because only galaxies with
magnitudes close to $M$ contribute in the derivation of $\alpha (M)$.
We present in Tables 3 and 4 $\alpha_{mn}$, where 
$-0.4 (\alpha_{mn} + 1)$ 
is the slope of $\log_{10} N$ 
at $M$ of the best fitting polynomial of order $m$ to
the $2n+1$ points, computed at 1 magnitude intervals, 
centered on $M$. 
The case $m=1$ corresponds to a linear fit.

The tables show that $-1.5 < \alpha < -1.2$ for
$-18 < M_B < -13$ and $-19 < M_R < -15$.  This is similar to
what has been observed in the Coma cluster in this
magnitude range
(Secker \& Harris 1996, Trentham 1997c).
The tables also show that $\alpha = -1$ is nowhere
a good fit
to the data: the LF is always rising.  
The LF is continuous
over the magnitude range where there is a 
discontinuity between giants and
dwarfs in the cluster fundamental plane (at $M_B \sim -16$).
This is suggestive of a conspiracy in which the giant galaxy
LF is falling by an amount 
almost exactly compensated
for by the rise in the dwarf LF, on going to
fainter magnitudes.  Such an effect was also seen in the
Virgo cluster (Sandage et al.~1985).

The pattern evident in Figure 5 is not visible in
Figures 3 or 4 because the statistics in each cluster are poor;
it is only when we combine all the data that the trend 
in Figure 5 becomes significant.
The reason that the statistics in each cluster are poor is
because the field-to-field variance in the background is high.
In a separate paper (Trentham 1997d), we extend this 
analysis to a much larger sample of clusters from our own data,
and from the literature.
We obtain a much more tightly constrained LF 
and argue that the LF in 
clusters
might be universal. 

\vskip 5pt

\noindent{\bf 3.3 Colours}

\noindent
In Figure 6, we present the colours of galaxies having absolute
magnitude $-16 < M_R < -15$
in the observed frame.  This narrow magnitude range was chosen as it 
encompasses all magnitudes bright enough to be reliably 
probed by the data we have for all four clusters, while still being
faint enough that the galaxies in this range are likely to be dwarf
galaxies as defined by their
fundamental plane positions 
(very few giant galaxies exist this faint; Binggeli 1987, 1994). 
The colours are computed as described in the figure caption and
in Section 2.2 (step 4) of the text.

The colour distribution in Abell 1367 is essentially unconstrained,
for the same reason that the LF is so poorly determined i.e.~the galaxy
density is too low relative to the background.   
The statistics for
Abell 2199 and Abell 1795 are much better (although still not  
very good).  
The mean colour (corrected for background contamination) of galaxies  
blueward of giant ellipticals at the cluster redshift 
is $B-R = 1.31$ for Abell 2199 and $B-R = 1.70$ for Abell 1795.   
This colour cut removes galaxies which are very likely to be
background galaxies
(although they could, in principle, be reddened cluster
members), and therefore improves the statistics.
For Abell 1146, the mean 
colour of galaxies in excess of background with $1.0<B-R<2.8$ 
is $B-R$ = 2.06 (here we impose a cut at the red end to remove 
background contamination, and at the blue end to assist in
removing possible stellar contamination $-$ see Figure 6).  
The mean colours, as computed above,
are presented as a function of redshift in Figure 7.
The figure suggests that in all cases 
here, and for Coma (where the statistics are much better), the galaxies
are too red to be dwarf irregulars, and are probably dwarf spheroidals
(see Trentham 1997b for a discussion of the colours of the different
kinds of galaxies). 

\vskip 10pt

\noindent{\bf ACKNOWLEDGMENTS}

\noindent
This research has made use of the NASA/IPAC extragalactic database (NED) which
is operated by the Jet Propulsion Laboratory, Caltech, under agreement with the
National Aeronautics and Space Administration.

\vskip 10pt

\ni{\bf REFERENCES }
\beginrefs

Abell G.~O., 1958, ApJS, 3, 211

Arnaud M., Rothenflug R., Boulade O., Vigroux L., Vangioni-Flam E., 
1992, A\&A, 254, 49

Bernstein G.~M., Nichol R.~C., Tyson J.~A., Ulmer M.~P., Wittman D., 1995, AJ,
110, 1507

Binggeli B., 1987, in Faber S.~M., ed., Nearly Normal Galaxies.
Springer-Verlag, New York, p.~195

Binggeli B., 1994, in Meylan G., Prugneil P., ed., ESO
Conference and Workshop Proceedings No.~49: Dwarf Galaxies. 
European Space Observatory, Munich, p.~13

Broadhurst T.~J., Ellis R.~S., Shanks T., 1988, MNRAS, 235, 827

Burstein D., Heiles C., 1982, AJ, 87, 1165
 
Cardelli J.~A., Clayton G.~C., Mathis J.~S., ApJ, 345, 245

Coleman G.~D., Wu C-C., Weedman D.~W., 1980, ApJS, 43, 393

Cowie L.~L., Hu E.~M., Songaila A., 1995, Nat, 377, 603

Cowie L.~L., Songalia A., Hu E.~M., 1991, Nat, 354, 460

Cowie L.~L., Songaila A, Hu E.~M., Cohen J.~G., 1996, AJ, in press 

De Propris R., Pritchet C.~J., Harris W.~E., McClure R.~D., 1995, ApJ, 450, 534

Driver S.~P., Phillipps S., Davies J.~I., Morgan I.,
Disney M.~J, 1994, MNRAS, 268, 393  

Fabian A.~C., 1994, ARAA, 32, 277

Ferguson H.~C., 1989, AJ, 98, 367 

Ferguson H.~C., 1993, MNRAS, 263, 343

Ferguson H.~C., Binggeli B., 1994, A\&AR, 6, 67

Ferguson H.~C, McGaugh S.~S., 1995, ApJ, 440, 470

Ferguson H.~C., Sandage A., 1991, AJ, 101, 765

Gioia I.~M., Luppino G.~A., 1994, ApJS, 94, 583

Girardi M., Fadda D., Giuricin G., Mardirossian F., Mezetti M.,
Biviano A., 1996, ApJ, 457, 61

Godwin J.~G., Peach J.~V., 1982, MNRAS, 200, 733

Hodge P.~W., 1989, ARAA, 27, 139 

Hughes J.~P., Gorenstein F., Fabricant D., 1988, ApJ, 329, 82

Jarvis J.~F., Tyson J.~A., 1981, AJ, 86, 476 

Jones C., Forman W., 1984, ApJ, 276, 38

Jones L.~R., Fong R., Shanks T., Ellis R.~S., Peterson B.~A., 1991,
MNRAS, 249, 481.

Kormendy J., 1985, ApJ, 295, 73

Kormendy J., 1987, in Faber S.~M. ed., Nearly Normal Galaxies. 
Springer-Verlag, New York, p.~163

Kormendy J., 1990, in Kron R.~G., ed., The Edwin Hubble Centennial Symposium:
The Evolution of the Universe of Galaxies.  Astronomical Society of the 
Pacific, San Francisco, p.~33

Kormendy J., Bender R., 1994, in Meylan G., Prugneil P., ed., ESO
Conference and Workshop Proceedings No.~49: Dwarf Galaxies.
European Space Observatory, Munich, p.~161

Landolt A.~U., 1992, AJ, 104, 340

McGaugh S.~S., 1994, Nature, 367, 538

Sandage A., Binggeli B., Tammann G.~A., 1985, AJ, 90, 1759

Schechter P., 1976, ApJ, 203, 297

Secker J., 1996, preprint

Secker J., Harris W.~E., 1996, ApJ, 469, 628

Trentham N., 1997a, MNRAS, in press

Trentham N., 1997b, MNRAS, submitted 

Trentham N., 1997c, MNRAS, submitted 

Trentham N., 1997d, MNRAS, submitted 

Valdes F., 1982, Proc.~SPIE, 331, 465 

Valdes F., 1989, in Grosbol P.~J., Murtagh F., Warmels R.~H., ed.,
ESO Conference and Workshop Proceedings No.~31:
Proceedings of the 1st ESO/St-ECF Data Analysis Workshop.
European Space Observatory, Munich,  p.~35

Wilson G., Smail I., Ellis R.~S., Couch W.~J., 1997, MNRAS, in press

Zabludoff A. I., Huchra J.P., Geller M. J., 1990, ApJS, 74, 1

\endrefs

\vskip 10pt
 
\ni {\bf FIGURE CAPTIONS}
\vskip 10pt
\ni {\bf Figure 1.~} 
The
$R$-band images of our fields.  North is up and east is to the left in
all images; the field sizes are given in Table 2.
The cD galaxy (NGC 6166) of A2199 is to the west of Field II, and
its outer halo is visible in part to the right of the image. 
The cD galaxies of A1795 and A1146 are the brightest galaxies in those
images.

\vskip 10pt
\ni {\bf Figure 2.~} 
The number count $-$ magnitude relations, computed as described in the
text.  For A2199, the data from both fields are included.  
The dashed lines in each figure are the mean background counts from
Trentham (1997a).

\vskip 10pt
\ni {\bf Figure 3.~}The $B$-band luminosity functions for the sample
clusters. 
Slopes corresponding to $\alpha = -2$ and $\alpha = -1$ are shown
in each panel.
 
\vskip 10pt
\ni {\bf Figure 4.~}The $R$-band luminosity functions for the sample
clusters. 
Slopes corresponding to $\alpha = -2$ and $\alpha = -1$ are shown
in each panel.

\vskip 10pt
\ni {\bf Figure 5.~}The composite luminosity function for the 4 clusters
in this survey.  Both the $B$-band (upper panel) and $R$-band (lower
panel) functions are shown.  These are the weighted average of
the individual luminosity functions shown in Figures 3 and 4, where
each cluster is normalized to have the same number of galaxies
brighter than $M_B = -16.5$ or $M_R = -17.5$.  The vertical scales 
are arbitrarily chosen to
correspond to that of a cluster with 1000 galaxies deg$^{-2}$ brighter
than $M_B = -16.5$ or $M_R = -17.5$ (this would be a typical
Abell richness 2 cluster). 
Slopes corresponding to $\alpha = -2$ and $\alpha = -1$ are shown
in each panel.

\vskip 10pt
\ni {\bf Figure 6.~}Colour 
histograms for the sample clusters.  Only galaxies whose
isophotal magnitudes satisty the condition $-16 < R - \mu < -15$
are included in these plots, where $R$ is the isophotal magnitude
and $\mu$ is the distance modulus  
listed in Section 3.1 of the text.
Only objects classified as ``galaxy''
in both $B$ and $R$ images are included, except 
for A1146, where we include
all objects classified as ``galaxy'' in just the $R$-band image (the
reddest objects appear very small and faint in the $B$-band image of
A1146, and
are subject to misclassification as ``stars'').
All colours are measured 
within a 3\sd0 diameter aperture centered on
the galaxy.  The histograms include a background subtraction,
using the background data presented in Trentham 1997a; 
the background measurements were made in
an identical way to the cluster measurements in each case.   
In some cases there were more galaxies in the background field
per square degree
than galaxies in the cluster
field in a given bin $-$ hence the negative values on
the ordinate axis.

Horizontal error bars are small.  Vertical error bars correspond to
the quadrature sum of the cluster field and background error bars.
Typically, the background counts in each bin are of the same order
of magnitude as the cluster counts, so that an estimate of the
vertical error in each bin is $\sqrt{2N}$ where $N$ is the number of
galaxies.  These are generally very large. 
The histograms are complete to at least $B-R \sim 2.7$ in all cases.

\vskip 10pt
\ni {\bf Figure 7.~}  
The mean colour of the excess galaxies detected above the background with
$-16 < R - \mu < -15$, as a function of the  
cluster redshift.  These numbers are computed as described in the
text.  The Coma point is from Secker (1996).  
Predicted curves for dIrr and giant ellipticals are presented;
these use the $K$-corrections of Coleman et al.~(1980). 
The shaded region is where we predict local dSphs to lie, given the
$K$-corrections of Trentham (1997b). 
The dSphs exhibit a range in colour,
because they exhibit a range in metallicity and star formation
history.

\par\vfill\eject\bye